\documentclass[twocolumn]{aastex63}
\usepackage{newtxtext,newtxmath}
\usepackage{threeparttable}
\usepackage{color}

\shorttitle{Quasar X-ray/UV Luminosity Correlation}
\shortauthors{Singal et al.}

\begin{document}

\title{THE X-RAY LUMINOSITY FUNCTION EVOLUTION OF QUASARS AND\\ THE CORRELATION  BETWEEN THE X-RAY AND ULTRAVIOLET LUMINOSITIES}

\author{J. Singal}

\affiliation{Physics Department, University of Richmond\\138 UR Drive, Richmond, VA 23173}
\affiliation{Also Visiting Scholar, Kavli Institute for Particle Astrophysics and Cosmology, Stanford University}

\author{S. Mutchnick}

\affiliation{Physics Department, University of Richmond\\138 UR Drive, Richmond, VA 23173}

\author{V. Petrosian}

\affiliation{Department of Physics and Kavli Institute for Particle Astrophysics and Cosmology, Stanford University\\382 Via Pueblo Mall, Stanford, CA 94305-4060}
\affiliation{Also Department of Applied Physics, Stanford University}

\email{jsingal@richmond.edu}

\begin{abstract}
We explore the evolution of the X-ray luminosity function of quasars and the intrinsic correlation between the X-ray and 2500\AA $\,$ ultraviolet luminosities, utilizing techniques verified in previous works and a sample of over 4000 quasars detected with Chandra and XMM-Newton in the range $0<z<5$. We find that quasars have undergone significantly less evolution with redshift in their total X-ray luminosity than in other wavebands.  We then determine that the best fit intrinsic power law correlation between the X-ray and ultraviolet luminosities, of the form $L'_{\rm X} \propto ({L'_{\rm UV}})^{\gamma}$, is $\gamma=0.28\pm$0.03, and we derive the luminosity function and density evolution in the X-ray band.  We discuss the implications of these results for models of quasar systems.   
\end{abstract}


\section{Introduction} \label{intro}

In active galactic nucleus (AGN) systems such as quasars, the accretion disk, jets, and other possible structures such as dusty tori and coronae, collectively glow in wavebands (e.g. optical, radio, infrared, ultraviolet, X-ray, gamma-ray, etc.) across the electromagnetic spectrum.  Various works have attempted to access the correlations between luminosities in widely separated wavebands in AGN systems \citep[e.g.][]{Dermer07,La10,QP2,QP3}, in part to better understand the physics of these systems.  

Recently \citet{RL} used a measure of the correlation between the X-ray and ultraviolet luminosities of quasars to arrive at a determination of the shape of the luminosity distance function in the range $1.4<z<5$.  The intrinsic correlation between the X-ray and ultraviolet luminosities of quasars is thus a quantity of renewed interest.  Previous works have also explored the question of this correlation \citep[e.g.][]{AT86,Just07}. 

\cite{CR1} outlined a procedure for obtaining the intrinsic correlation between optical and radio luminosity. In a more recent work \citep{CW} we explored in depth the question of to what extent observed correlations in multiwavelength flux-limited data are indicative (or not) of intrinsic correlations.  As highlighted there, observed correlations can be quantitatively and qualitatively very different from the intrinsic correlations due to multi-dimensional observational selection effects which truncate the data, the redshift evolution of the luminosities, the common dependence on the luminosity distance (obtained from redshift for a given cosmological model).

Here we present an independent determination of the intrinsic correlation between the X-ray and ultraviolet (UV) luminosities of quasars, using the techniques verified in \citet{CW} and applied to other wavebands in quasars \citep{QP1,QP2,CR1,QP3} and blazars \citep{BP1,BP2,BP3,V21}.  We utilize two joint X-ray-ultraviolet data sets, based on the overlap of X-ray observations from the Chandra and XMM-Newton X-ray observatories, respectively, with the quasar catalog of the Sloan Digital Sky Survey (SDSS) for which the rest frame 2500\AA$\,$ UV fluxes have been provided.  In \S \ref{data} we discuss the data sets and in \S \ref{tech} we extract the intrinsic redshift evolution of the X-ray luminosity and the X-ray-UV luminosity-luminosity correlation.  We derive the local X-ray luminosity function in \S \ref{lumsec}.  \S \ref{disc} presents a discussion, including the implications of the luminosity evolution and the correlation between these luminosities for our understanding of quasar systems.

\begin{figure}
\includegraphics[width=3.5in, height=2.5in]{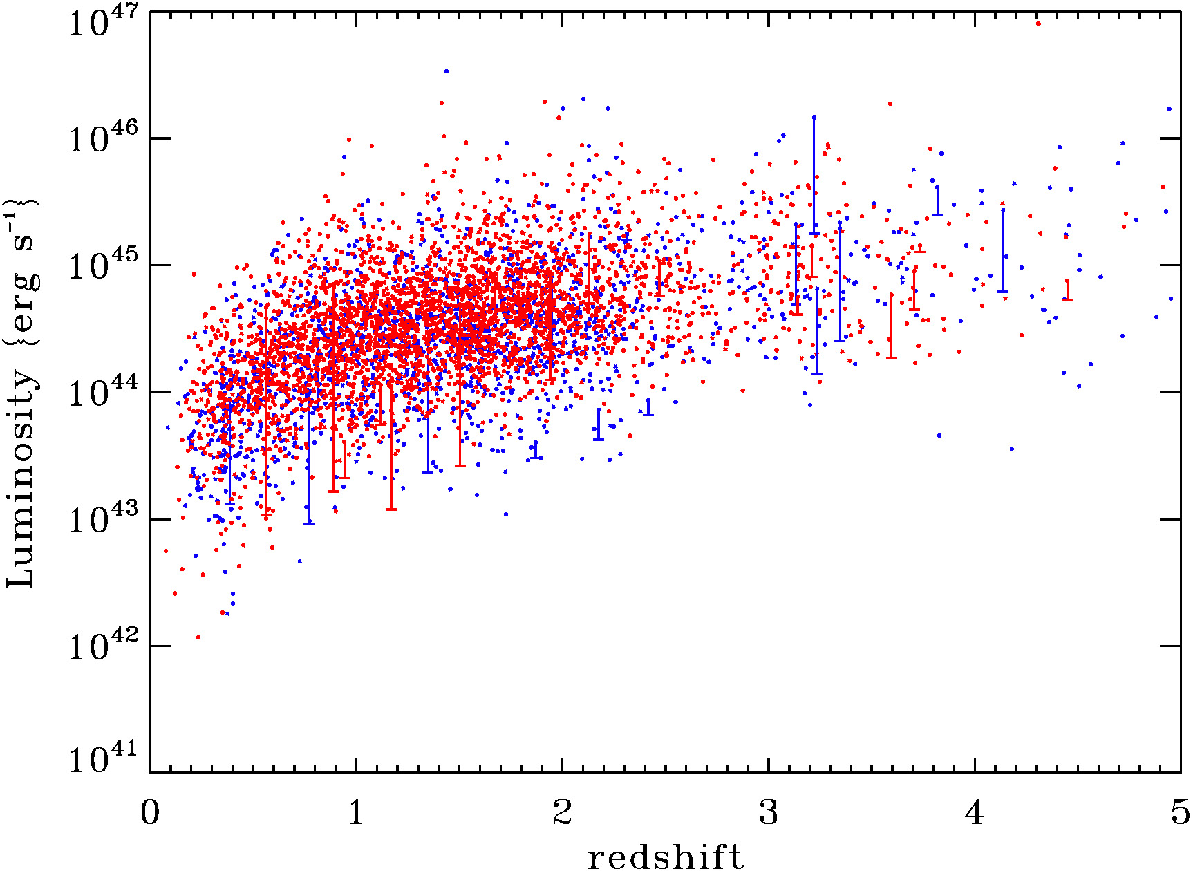}
\caption{The X-ray luminosities versus redshift for the Chandra (blue) and XMM-Newton (red) quasar data sets.  For a few objects (but not all, for clarity) we show the limiting X-ray luminosities that the object could have for inclusion in the data set given its redshift, discussed in \S \ref{data}. }
\label{xlums}
\end{figure} 

\begin{figure}
\includegraphics[width=3.5in, height=2.5in]{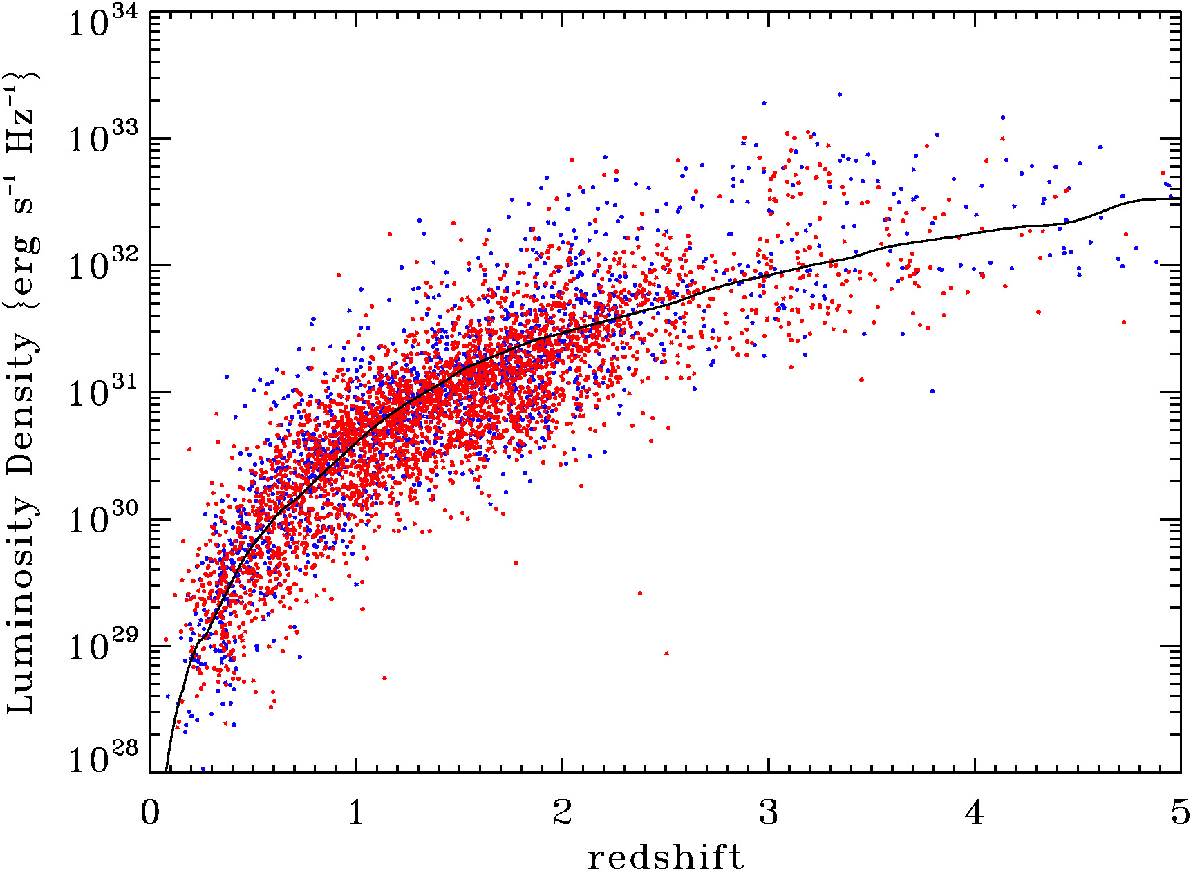}
\caption{The 2500\AA$\,$ monochromatic UV luminosity density versus redshift for the Chandra (blue) and XMM-Newton (red) quasar data sets.  The luminosities are calculated from the rest frame 2500\AA$\,$ fluxes presented by \citet{Shen11} using Equation 9\ref{Luv}).  An illustration of the limiting luminosity (including full emission line and continuum K-corrections) corresponding to $i$-band magnitude $\leq$ 19.1 to create an optically flux-limited data set as discussed in \S \ref{data} is also plotted, assuming a spectral index of $\varepsilon_{UV}=0.5$ to convert from $i$-band to 2500\AA.}
\label{uvlums}
\end{figure} 

\nopagebreak \section{Data Sets} \label{data} 
\nopagebreak A large catalog of identified quasars with spectroscopic redshifts is provided by the SDSS.  We use the well-established SDSS data release 7 (DR7) quasar catalog \citep{SDSSQ}, which contains over 105,000 objects.  In the SDSS DR7, objects were identified as quasar candidates for spectroscopic follow-up if they displayed the requisite optical colors, or if identified via ROSAT X-ray data, or if selected by a  so-called ``serendipity'' algorithm that identifies unusual colors in concert with a radio match, or if they had a radio match within $2''$.  For parts of our analysis as discussed below in \S \ref{techLE} and \S \ref{lumsec} we  select a sample of SDSS quasars with a well defined flux limit by imposing an $i$-band magnitude upper limit of 19.1, as  in our previous works \citep{QP2,QP3}.  As discussed in \citet{SDSSQ} and \citet{R06} this results in a catalog with a smoother redshift distribution with a reduced bias toward objects with $z>$2.  We use UV rest frame monochromatic fluxes at 2500\AA$\,$, $F_{2500}$, taking into account spectral shapes and emission line effects provided by \citet{Shen11} for all DR7 quasars, to obtain the UV monochromatic luminosity density

\begin{equation}
\label{Luv}
L_{2500}=4\pi \, D_L(z) \, F_{2500}
\end{equation}

We utilize the overlap of the DR7 quasar catalog with observations in the Chandra X-ray observatory general source catalog \citep{Chandra} release 2 and the fourth XMM-Newton X-ray observatory serendipitous source catalog \citep{XMM} release 9, with a $3''$ matching criterion.  If a DR7 quasar was present in both X-ray catalogs we considered only the Chandra observation so the two data sets feature all unique objects.  This results in 1592 and 3193 unique sources, respectively, after applying the likelihood cuts discussed below.  

Both X-ray catalogs present a measured wide-band aperture energy flux for each object as well as the aperture energy flux in several sub-bands if available, as summarized in Table \ref{tab1}.  Most sources in these catalogs are missing values for the fluxes in one or more sub-bands, so in this analysis we use the broad band ($0.5-7$ keV for Chandra and $0.2-12$ keV for XMM-Newton) fluxes as the determinative X-ray flux for each object.  The sub-band fluxes are used only for the determination of each object's X-ray spectral index as discussed below.  There are 4 and 5 sub-bands, respectively, for the two instruments, with similar widths and central values shown in Table \ref{tab1}.  The small difference between the broad bands of the two instruments can be ignored for our analysis so we can treat X-ray fluxes  to be centered at roughly at 2 keV.

\begin{table}
\scriptsize
\caption{X-ray energy bands utilized }
\label{tab1}
\begin{center}
\begin{tabular}{rrrr}
 & \textbf{Chandra}& \textbf{XMM-} & \textbf{Newton} \\
name & energy & name & energy \\
 & (keV) &  & (keV)\\
\hline
``b'' (broad) & 0.5-7 & ``total'' (broad) & 0.2-12\\
``u'' & 0.2-0.5 & ``1'' & 0.2-0.5\\
``s'' & 0.5-1.2 & ``2'' & 0.5-1.0\\
``m'' & 1.2-2.0 & ``3'' & 1.0-2.0\\
``h'' & 2.0-10 & ``4'' & 2.0-4.5\\
 &   & ``5'' & 4.5-12\\
\hline
\end{tabular}
\begin{tablenotes}
\footnotesize
\item As discussed in \S \ref{data} in this analysis we use the broad band fluxes as the determinative X-ray flux, $F_X$, for each object.  The sub-band fluxes are used only for the determination of each object's X-ray spectral index.
\end{tablenotes}
\end{center}
\end{table}

From the broad band fluxes we obtain X-ray luminosity as:
\begin{equation}
\label{Lxray}
L_X={{4\pi \, D_L(z) \, F_X} \over {K_X(z)}},
\end{equation}
where for the $K$ correction, $K_X(z)$, we assume a power-law spectrum 
\begin{equation}
L_X \propto \nu^{-\varepsilon_X}
\label{spectpow}
\end{equation}
so that the K-correction is then
\begin{equation}
K_X(z) = (1+z)^{1-\varepsilon_X}.
\end{equation}
The scatter diagrams of X-ray and UV luminosity versus redshift of the objects in the two data sets are shown in Figure \ref{xlums} and  Figure \ref{uvlums}, respectively.

\begin{figure}
\includegraphics[width=3.5in, height=2.5in]{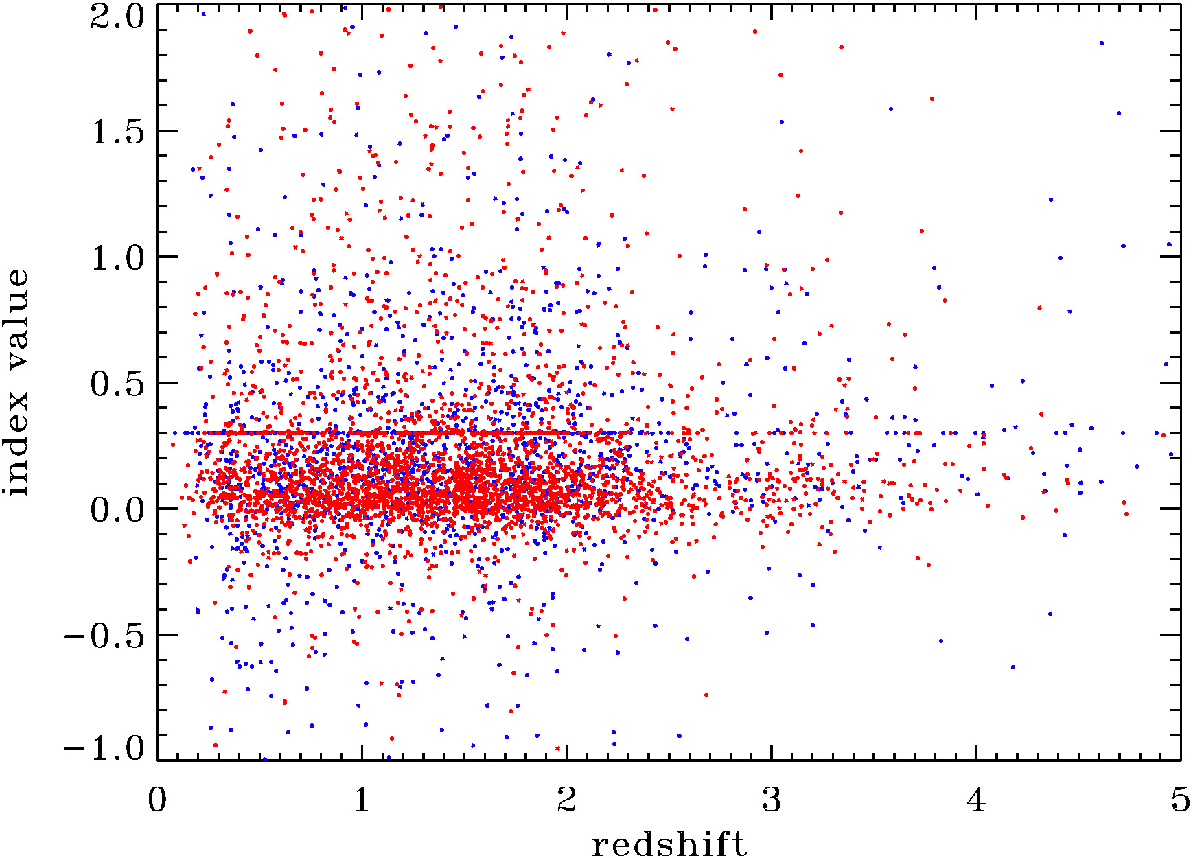}
\caption{Best-fit X-ray spectral index values $\varepsilon_x$ versus redshift for the Chandra (blue) and XMM-Newton (red) quasar data sets, determined from the reported sub-band energy fluxes as discussed in \S \ref{data}.  }
\label{xks}
\end{figure}

For the X-ray band, we individually fit the spectral index $\varepsilon_x$ for each object whenever possible, utilizing the energy flux values reported in the sub-bands shown in Table \ref{tab1}.  For a minority of objects the number of sub-bands with an energy flux value reported is too small and/or the flux values themselves are too discordant so as to produce a fitted spectral index outside of the range from $-1\leq \varepsilon_x \leq 2$ and thus which we consider unlikely to be realistic.  This is the case for only 152 objects in the Chandra data set and 200 objects in the XMM-Newton data set, and for these objects we assign a spectral index equal to 0.3, which is a compromise between the mean index of the fitted objects of 0.2 and values of $\varepsilon_x$=0.4--0.7 often assumed for X-ray AGNs \citep[e.g][]{La10}, noting again that this applies to only $<10\%$ of objects in our data sets and that the results of this analysis are not particularly sensitive to small changes in $\varepsilon_x$.  Figure \ref{xks} shows the fitted X-ray spectral indexes.

The X-ray catalogs do not have a single flux limit; rather, the limiting flux for the inclusion of any given object is a function of the parameters of its observation, including the background surface brightness at its location and the duration of observing.  As part of our analysis discussed in \S \ref{tech} requires knowledge of an object's theoretical lowest flux for inclusion in the data set, we calculate the limiting flux for inclusion of any given object as follows:  Each object has an assigned detection likelihood value, which can be converted to an equivalent detection significance for the broad band by fitting an exponential function to a detection significance vs. likelihood plot provided in the explanatory materials for the catalogs.\footnote{For example, Chandra's is provided at https://cxc.harvard.edu/csc/columns/significance.html}.  
We also assign a minimum likelihood value for an object's inclusion in our data sets, based on recommendations in the explanatory materials for the catalogs, and eliminate objects with a lower likelihood value.  For example, \citet{XMM} recommend to set a minimum likelihood of 14 for the XMM-Newton sources to ensure that statistically almost all sources are real, while for Chandra this value is 8.  Applying the likelihood to significance conversion to these minimum likelihood values gives a minimum significance for inclusion in the data set.  The ratio of an object $j$'s significance $\sigma_j$ to its minimum significance ($\sigma_{j,min}$) can be used to calculate its minimum X-ray flux for inclusion in the data set with
\begin{equation}
F_{j,x,lim} = F_{j,x} \times { {\sigma_{j,min}} \over {\sigma_j}}
\end{equation}
The minimum X-ray luminosity that an object could have to be included in the survey at any redshift, utilized below in \S \ref{tech}, is a function of its lower limiting flux and redshift, with the standard luminosity-redshift relation of equations (\ref{Luv}):
\begin{equation}
L_{j,x,lim}(z) = {{ 4 \pi \, D_L\!(z)^2 \, {F_{j,x,lim} } } \over {K\!(z)} }
\label{lumlimeq}
\end{equation}
A similar procedure was used to deal with infrared fluxes and luminosities and their limits in a previous analysis \citep{QP3}.  Some minimum limiting X-ray luminosities are shown in Figure \ref{xlums}.  

\section{Luminosity Evolutions and Correlations} \label{tech}

As discussed at length both analytically and empirically in \citet{CW}, in order to access the true intrinsic correlation between luminosities in widely separated wavebands in a population of extragalactic objects distributed over a wide range of redshifts, one must correct the luminosities for this effect as originally described by \citet{EP92}, then  determine the nature of the intrinsic luminosity-luminosity correlation.

\subsection{Luminosity Evolutions} \label{techLE}

As pointed out by \citet{P92}, determination of the luminosity evolution is a critical step in the investigation of cosmological evolution of astrophysical sources. Presence of luminosity evolution implies a correlation between luminosity and distance or redshift. A common non-parametric method for determining the presence of a correlation is Spearman Ranking using the test statistic

\begin{equation}
\tau = {{\sum_{j}{(\mathcal{R}_j-\mathcal{E}_j)}} \over {\sqrt{\sum_j{\mathcal{V}_j}}}}
\label{tauen}
\end{equation}
that tests the independence of two variables, say ($x_j,y_j$) in a data set with $j=1,\dots, N$.  Here $\mathcal{R}_j$ is the dependent variable ($y$) rank of the data point $j$ in a subset, called the  {\it associated set}  consisting of $n$ sources. For independent variables  $\mathcal{E}_j=(1/2)(n+1)$ is the expectation value and $\mathcal{V}_j=(1/12)(n^{2}+1)$ is the variance. 

For untruncated data (i.e. data truncated parallel to the axes) the associated set  of point $j$ includes all of the points with a lower (or higher, but not both) independent variable value ($x_k < x_j$).  \citet{EP92} developed  non-parametric methods for determination of the correlation for generally truncated data, which we have applied in previous works and verified most recently with simulations in \citet{CW}.  The truncation modifies the associated set to be a sub set consisting only of those points of lower (or higher, but not both) independent variable ($x$) value {\it that would have been observed if they were at the $x$ values $\geq x_j$, or had $y$ values $\geq y_{\rm min, j}$  given by the truncation}. For a simply one-sided truncation, e.g.~a sample with well defined universal truncation, Figure 7 of \citet{CW} shows a graphical description of an associated set for a given example data point.  For the X-ray data, with each source having unique flux limit alone the associated set for object $k$ would be all objects whose luminosity is greater than their minimum limiting luminosity calculated at the redshift of object $k$ (i.e. equation \ref{lumlimeq} with $z=z_k$).  
The ultraviolet data is untruncated as every object in the SDSS DR7 quasar catalog has an ultraviolet flux reported by \citet{Shen11}.

For problem at hand, if the variables $z$ and $L$ are independent (i.e.~no luminosity evolution) then the ranks $\mathcal{R}_j$ should be distributed randomly and $\tau$ should sum to near zero.  Independence is rejected at the $m \, \sigma$ level if $\vert \, \tau \, \vert > m$.  To find the best form for correlation between $L$ and $z$, in other words for luminosity evolution, we define a new variable, $L'_j=L_j/g(z_j)$,  and the rank test is repeated, with different values of parameters of the function $F$ until $L'$ and $z$ are determined to be uncorrelated. Traditionally, ever since discovery of evolution of quasars by \citet{Schmidt67}, the evolution function $g(z)=(1+z)^k$ has been commonly used. However, such a form implies very rapid  evolution  during  the short age of the universe  at high redshifts. In our past application of the Efron-Petrosian method to AGNs \citep[e.g.][]{QP2,QP3}, and GRBs, extending to redshifts $z\gg 2$ we have found that the functional form
\begin{equation}
g_a(z)={(1+z)^{\rm k_a} \over 1+ \left( {{1+z} \over {Z_{\rm cr}}} \right)^{\rm k_a}}
\label{eveq2}
\end{equation}
with a value of $Z_{\rm cr}=3.7$ reduces the rate of the evolution at redshifts $z>Z_{\rm cr}-1$ and well describes the luminosity evolution in several wavebands over a large redshift range in real quasar data.  The task then is to find the evolution exponent $k_a$ for any given waveband, which renders each luminosity $L'$ (the dependent variable) uncorrelated with redshift (the independent variable). The luminosities thus obtained are referred to as the {\bf ``local luminosity"}.

\begin{figure}
\includegraphics[width=3.5in]{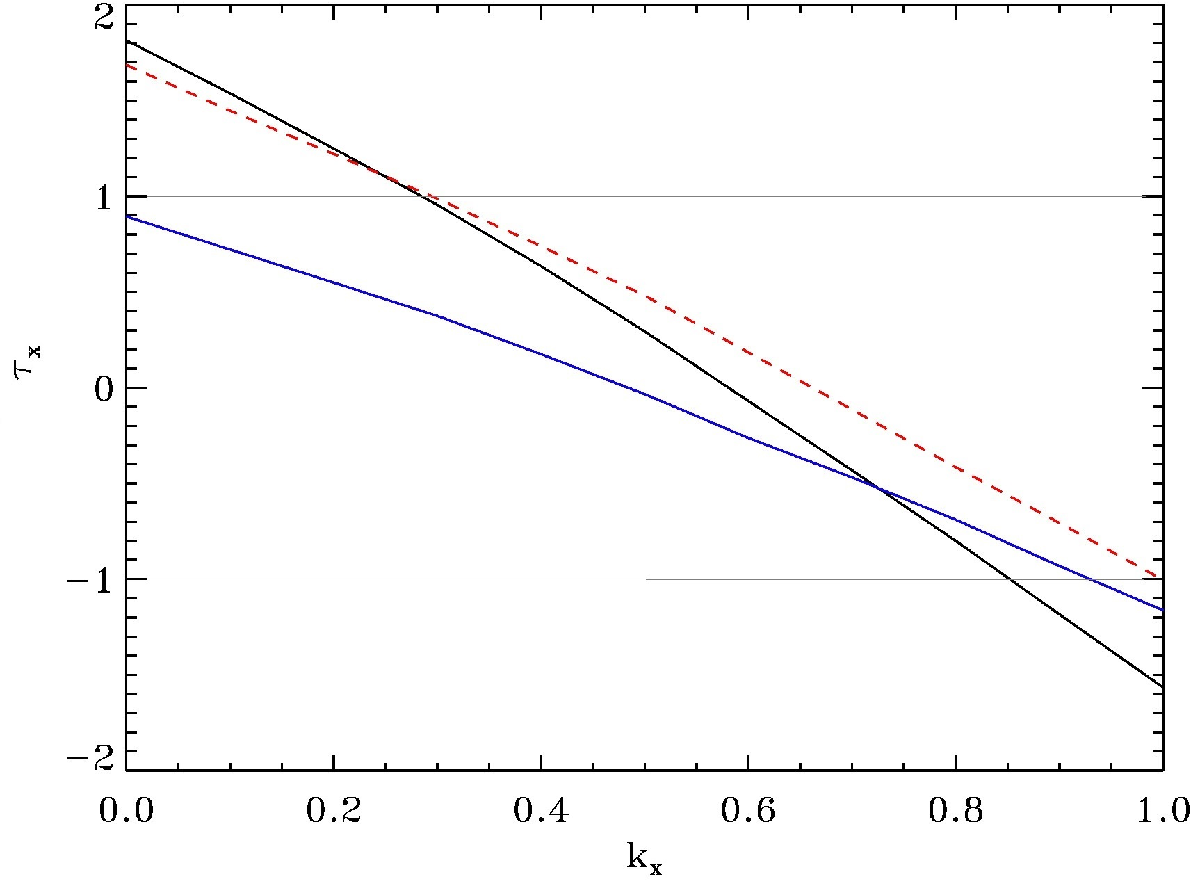}
\caption{$\tau_{\rm x}$ vs $k_{\rm X}$ for the combined (black), Chandra (blue), and XMM-Newton (red) data sets, for the forms of the luminosity evolutions given by equation \ref{eveq2}.  The 1$\sigma$ range of the best-fit $k_{\rm X}$ is where -1 < $\tau_{\rm x}$ < 1.}
\label{alphas}
\end{figure} 

Since optical observation is needed to  determine the spectroscopic redshift, optical flux, luminosity, and luminosity become relevant.  Objects must have sufficient flux in both optical (SDSS $i$-band) and X-ray bands to be included in these data sets.  Because we have two criteria for truncation, the associated set for each object $k$ includes only those objects that are sufficiently luminous in both X-ray and optical bands to have been in the survey if they were located at the redshift of the object in question.  The luminosity cutoff limits for a given redshift must also be adjusted by factors of $g_{\rm x}\!(z)$ and $g_{\rm opt}\!(z)$. 

However the multiband situation is somewhat alleviated in the present situation because previous works \citep{QP1,QP2,QP3} have converged on a robust determination of $k_{\rm opt}$=3.3 for DR7 quasars whether treating a large optical-only data set or simultaneously determining $k_{\rm opt}$ along with the best-fit evolution in some other band.  We assume that $k_{\rm opt}$ and $k_{\rm UV}$ will be the same given that they are essentially the same waveband (and indeed in previous analysis we scaled all optical luminosities to 2500 \AA).  Therefore we fix $k_{\rm UV}=k_{\rm opt}=3.3$  and determine the best-fit $k_{\rm x}$.  Figure \ref{alphas} shows the results for the test statistic $\tau_{\rm x}$ versus $k_{\rm x}$, with the best-fit $k_{\rm x}$ value being where $\tau_{\rm x}=0$ and the 1 $\sigma$ range obtained where -1 < $\tau_{\rm x}$ < 1. Results for the two data sets are in agreement so the data sets can be combined for tighter statistical uncertainty, resulting in a best-fit value of $k_{\rm x}$=0.55$\pm$0.30.  This is significantly lower (less evolution with redshift) than the values achieved for other wavebands.   We examine this issue in \S \ref{disc}.

\subsection{Luminosity-Luminosity Correlation} \label{techcorr}

\begin{figure}
\hspace*{-0.1in} 
\includegraphics[width=3.5in]{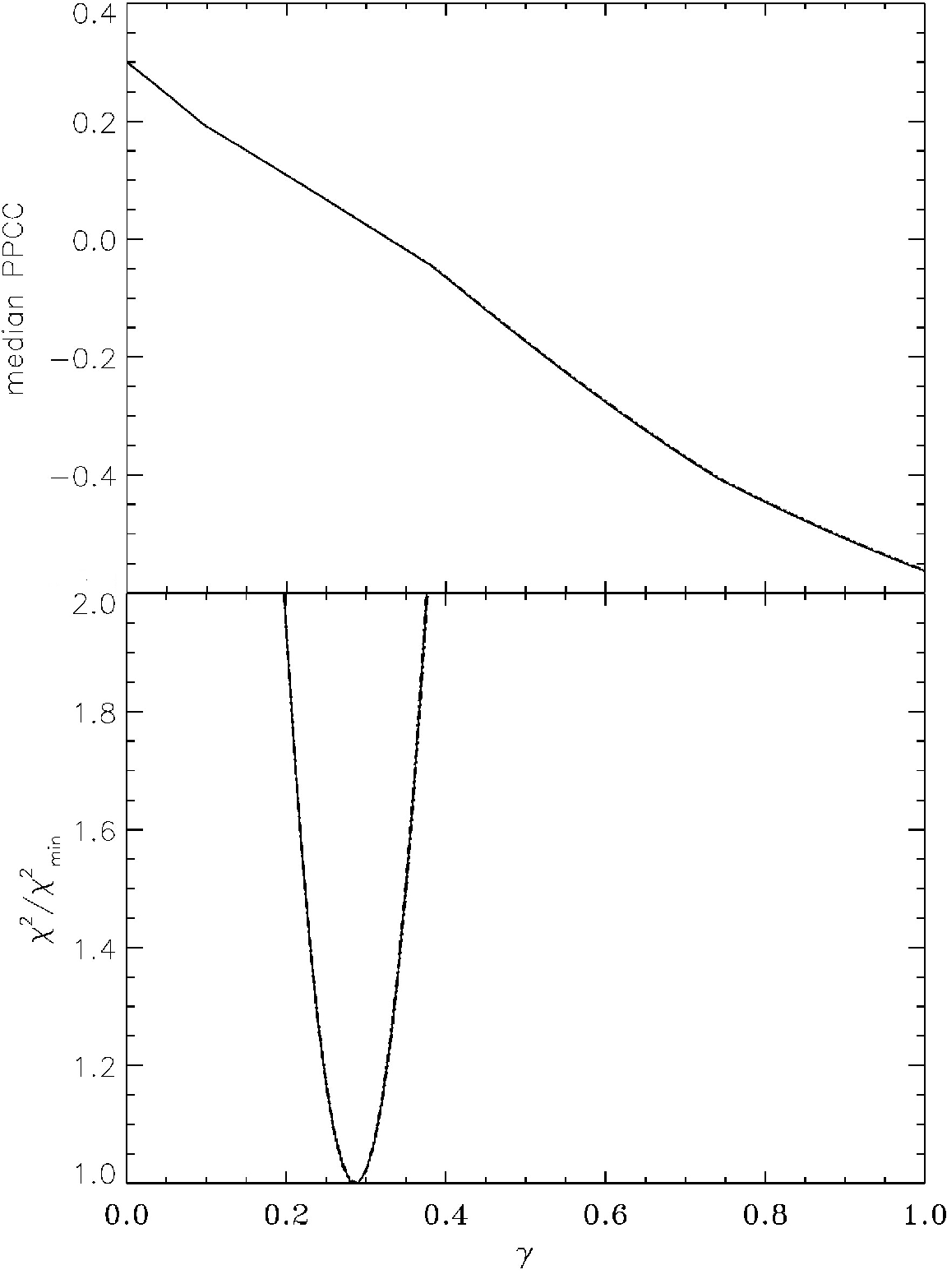}
\caption{Median (top) and total squared residual (bottom) of the binned PPCC values for the ``correlation reduced local luminosity'' (see equation \ref {deltaeq}).  As described in \S \ref{techcorr}, a median PPCC value of zero gives the best-fit luminosity-luminosity correlation power law index  $\gamma$, in this case $\gamma$=0.28$\pm$0.03.  Results are shown  for the best-fit value of $k_{\rm x}$ (solid curve) and the extremal 1$\sigma$ values of $k_{\rm x}$ (dash-dot curves), which sit nearly on top the solid curve.  As demonstrated by simulations in \citet{CW} this technique is reliable for recovering the power-law form of the intrinsic correlation. 
}
\label{deltas}
\end{figure}

For determination of the intrinsic correlation between the X-ray and ultraviolet luminosities, we use the procedure described in \citet{CR1}, which was verified analytically and by simulation in \citet{CW}, as follows:  

A standard measure of correlations between two variables 1 and 2, in our case luminosities, mutually dependent on a 3rd independent variable, here redshift, is the Pearson partial correlation coefficient \citep[PPCC --- e.g][]{RS07}: 
\begin{equation}
r_{12,3} = {{r_{12} - r_{13}r_{23}} \over {[(1-r_{13}^{2})(1-r_{23}^{2})]^{1/2}}},
\label{pearsoneq}
\end{equation}
where $r_{ab}$ is the standard sample Pearson's moment correlation (PMC  --- commonly known as Pearson's $r$) between variables $a$ and $b$;
\begin{equation} \label{pmc}
r_{ab} = \frac{\sum_{i} (a_{i}- \overline{a}) (b_{i}- \overline{b})}{N \sigma_a \sigma_b},
\end{equation}
where $\sigma_a = \sum \sqrt{\frac{1}{N}(a_{i}- \overline{a})^2}$ is the standard deviation of the $a$ values and $N$ is the total number of data points.  Because of the rapid decrease of number of sources with increasing luminosity, we calculate PMCs and PPCCs using the {\it logarithm} of the luminosity values (and linear redshift values), in order to reduce the potential outsize effect of a small number of objects with a very high luminosity. 

It is important to note that the PMC and PPCC are measures of the extent to which two variables are correlated, in the sense of being related by some function.  However, they do not shed any light on the nature of the correlation function itself, and a higher value does not necessarily indicate a steeper correlation function (or vice-versa). They only indicate that the data more closely adhere to the functional form whatever it may be.   

As mentioned above, in \citet{CW} using several sets of  simulated data  with different forms of intrinsic luminosity-luminosity correlations, we showed that that the binned PPCC between two local luminosities reliably reproduces the input parameters of the simulated correlation. In general the extant data allows a robust estimation of a  single-parameter correlation function. Following our earlier works \citep[e.g.][]{CR1,CW} we use a power law form for the luminosity-luminosity correlation (i.e.~a linear log-log correlation) by defining the so-called ``correlation reduced'' local X-ray luminosity $L'_{\rm crx}$ as

\begin{equation}
L'_{\rm crx} = {  {L'_{x} } \over { \left( {{ L'_{\rm UV} } \over {L_{\rm fid}} } \right)^{\gamma} } }
\label{deltaeq}
\end{equation}  
where $L_{\rm fid}$ is some fiducial luminosity, introduced to avoid exponentiating a dimensioned number.  Then for a range of values of $\gamma$ we compute the median value of the PPCC between $L'_{\rm crx}$ and $L'_{\rm UV}$ in bins.  The value of $\gamma$ that results in a median PPCC of zero is the best-fit value for the power law exponent for the intrinsic correlation between $L'_{X}$ and $L'_{\rm UV}$.  Figure \ref{deltas} shows the median PPCC vs. $\gamma$ and the sum of the squares of the residual values of the PPCCs  vs. $\gamma$.   The 1$\sigma$ range of uncertainties reported for these values is determined by considering the $\chi^2$ vs. $\gamma$ distribution.  We find that $\gamma$ is very insensitive to the value of $k_{\rm x}$ within the 1$\sigma$ range of the latter -- indeed the the PPCCs  vs. $\gamma$ curves 
corresponding to the extremal ends of the 1$\sigma$ range of $k_{\rm x}$, lie nearly on top of the curve corresponding to the middle value of $k_{\rm x}$=0.55 shown in Figure \ref{deltas}.  By considering the width of the sum of the residual values curve we obtain a best fit $\gamma$ value of 0.28$\pm$0.03. 

This completes the primary emphasis of this work. However, as we have done in other wavebands \citep[e.g.][]{QP2,QP3,CW}, below we present a complete   description of the cosmological distributions of the X-ray characteristics of quasars.

\section{Density evolution and Local X-ray Luminosity Function} \label{lumsec}

Following recent works \citep[e.g.][]{CW}, without loss of generality, the bi-variate luminosity function (LF) in some waveband $a$ as 
can be written as:
\begin{equation}
\Psi_{\rm a}(L_{\rm a},z) = \rho(z)\,{{\psi_{\rm a}\!(L_{\rm a}/g_{\rm a}(z) , \eta_{\rm j,a})} \over {g_{\rm a}(z)}},
\label{lumeq}
\end{equation}
where $g(z)$ and $\rho(z)$ describe the luminosity and (comoving) density evolutions, respectively, and $\eta_j$ stands for parameters that describe the shape (e.g. power law indices and break values) of the LF.  In what follows we assume a non-evolving shape for the LF (i.e. $\eta_{\rm j,a}= const$, independent of $L$ and $z$), which is a good approximation for determining the global evolution.  Allowing the shape of the LF to evolve would be a model featuring luminosity-dependent density evolution (LDDE) which has been shown to be a good model for the X-ray LF of AGN systems \citep[e.g.][]{Miyaji00,Ueda03,Ueda14,Miyaji15,Ftp16}.  However in our work we are seeking to determine characteristics of the LF with verified non-parametric methods, and LDDE has the disadvantage of not being able to be accessed with these techniques. The general representation of a LF given by equation \ref{lumeq} with $\eta_{\rm j,a}= const$ is fully adequate to determine the global bulk evolution factors and correlations between the luminosities which is the major emphasis of this work.

Having determined the luminosity evolution $g_{\rm X}(z)$ and established the local luminosity, $L'_{\rm X}=L_{\rm X}/g_{\rm X}(z)$, and redshift, $z$, as independent variables, we can obtain their mono-variate distributions, namely the  ``local'' LF $\psi_{\rm a}(L'_{\rm a})$ and density evolution $\rho(z)$ using the methods described below.

The quasar density evolution and LFs at radio, optical, infrared, and gamma-ray regimes were described in  \citet{QP1}, \citet{QP2}, \citet{QP3} and \citet{V21}.   Here we  focus on the X-ray regime.  As shown in those works, and also with simulations in \citet{CW}, the LFss in widely separated wavebands (for example optical and X-ray) are effectively separable -- i.e. the results determined when transforming to a pair of variables which are independent by taking out the best-fit correlation from one of the luminosities are very similar to those determined assuming the LFs in a given band and optical-UV band to be independent.

Having determined $g_{\rm x}\!(z)$ in \S \ref{techLE} we can determine the comoving density evolution $\rho(z)$ and local LF $\Phi(L'_{\rm x})$.  One can define the cumulative density function.  As demonstrated in \citet{P92}, the best non-parametric method of determining the distributions of independent variables in our case $\rho(z)$ and $\psi(L^\prime$) from a truncated data set is based on the $C^-$ method of \citet{L-B71}. This method gives a maximum likelihood estimate of the {\it cumulative} distributions of the variables; in our case the cumulative density evolution
\begin{equation}
\sigma(<z) = \int_0^z { {{dV} \over {dz}} \, \rho\!(z) \, dz},
\end{equation}
as
\begin{equation}
\sigma(<z) = \prod_{j}{\left[1 + {1 \over M(j)}\right]},
\label{sigmaeqn}
\end{equation}
and the cumulative local LF
\begin{equation}
\Phi(>L') = \int_{L'}^{\infty} {\psi(L'') \, dL''}
\end{equation}
as
\begin{equation}
\Phi(>L') = \prod_{k}{\left[1 + {1 \over N(k)}\right]}.
\label{phieq}
\end{equation}
Here  $M(j)$ and $N(k)$ are  the number of sources in source j's associated sets: $M(j)$ includes sources with $z_i<z_j$, and for our purpose, in order to account for the effects of the truncation due to the flux limits, the sources with sufficient optical and X-ray luminosity that they would be in the observed data set if they were at (the higher) redshift $z_j$. Similarly $N(k)$ include sources with local X-ray luminosities $L'_{{\rm x},i}>L'_{{\rm x},k}$ which would be in the survey if they were at object $k$'s luminosity considering the luminosity limits for inclusion in both optical and X-ray data sets.\footnote{Note that we build the cumulative density and local LFs from below and above, respectively.}

We then fit the cumulative functions $\sigma\!(z)$ and $\Phi(L'_{\rm x})$ by piecewise cubic spline functions and differentiate them locally at specific points to obtain the comoving density evolution $\rho(z)$ as:   
\begin{equation}
\rho\!(z) = {d \sigma\!(z) \over dz} \times {1 \over dV/dz}
\label{rhoeqn}
\end{equation}
and the differential local LF
\begin{equation}
\psi_{\rm x}\!(L_{\rm x}') = - {d \Phi_{\rm x}\!(L_{\rm x}') \over dL_{\rm x}'}
\label{psieqn}
\end{equation}

\begin{figure}
\includegraphics[width=3.5in]{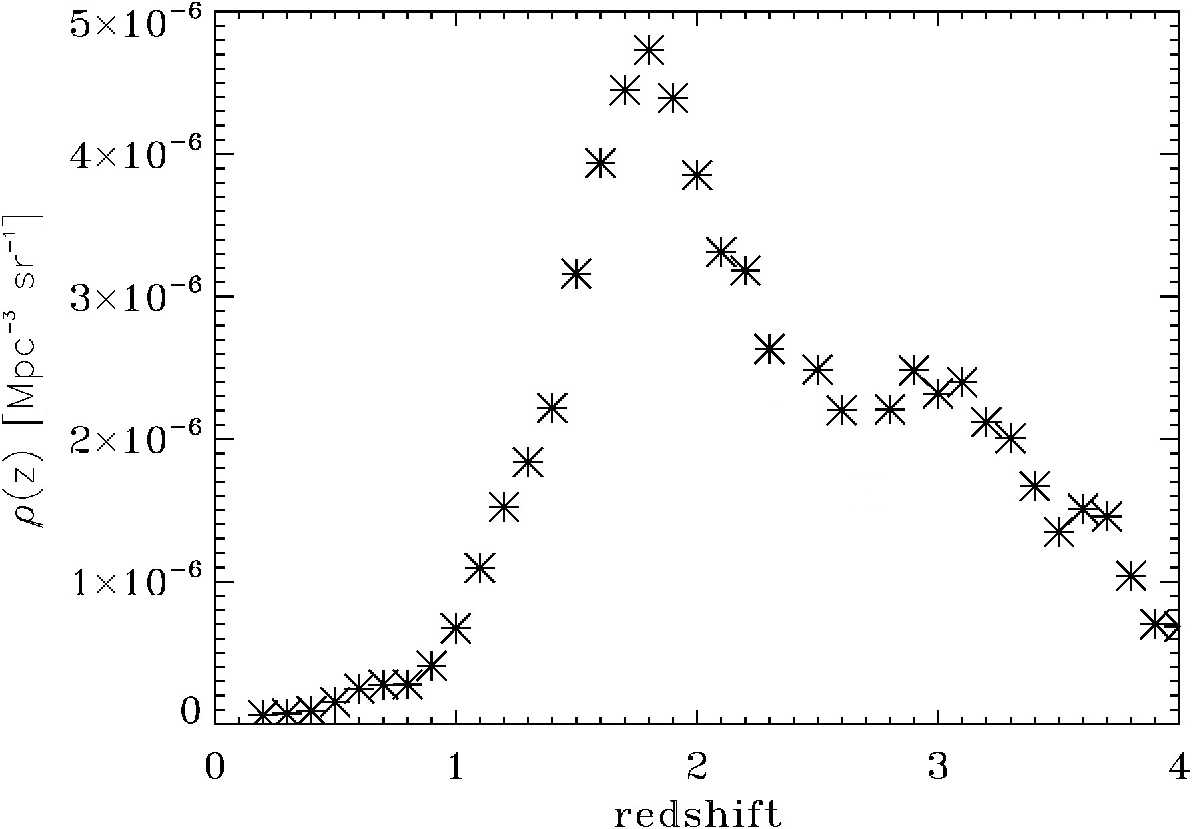}
\caption{The comoving density evolution $\rho(z)$ vs. redshift (large stars) for the for the quasars, calculated from the sample in this work as described in \S \ref{lumsec}.  $\rho(z)$ is defined such that $\sigma(z)=\int_0^{\infty} \rho(z) \, dV/dz \, dz$.  }
\label{rholog}
\end{figure}

Figure \ref{rholog} shows the density evolution function $\rho(z)$. It is seen that the density sharply peaks very near redshift 2, which is very similar to that obtained by other works which have analyzed the X-ray LF of AGN \citep[e.g.][]{Aird15,Ueda03}.  

\begin{figure}
\includegraphics[width=3.5in]{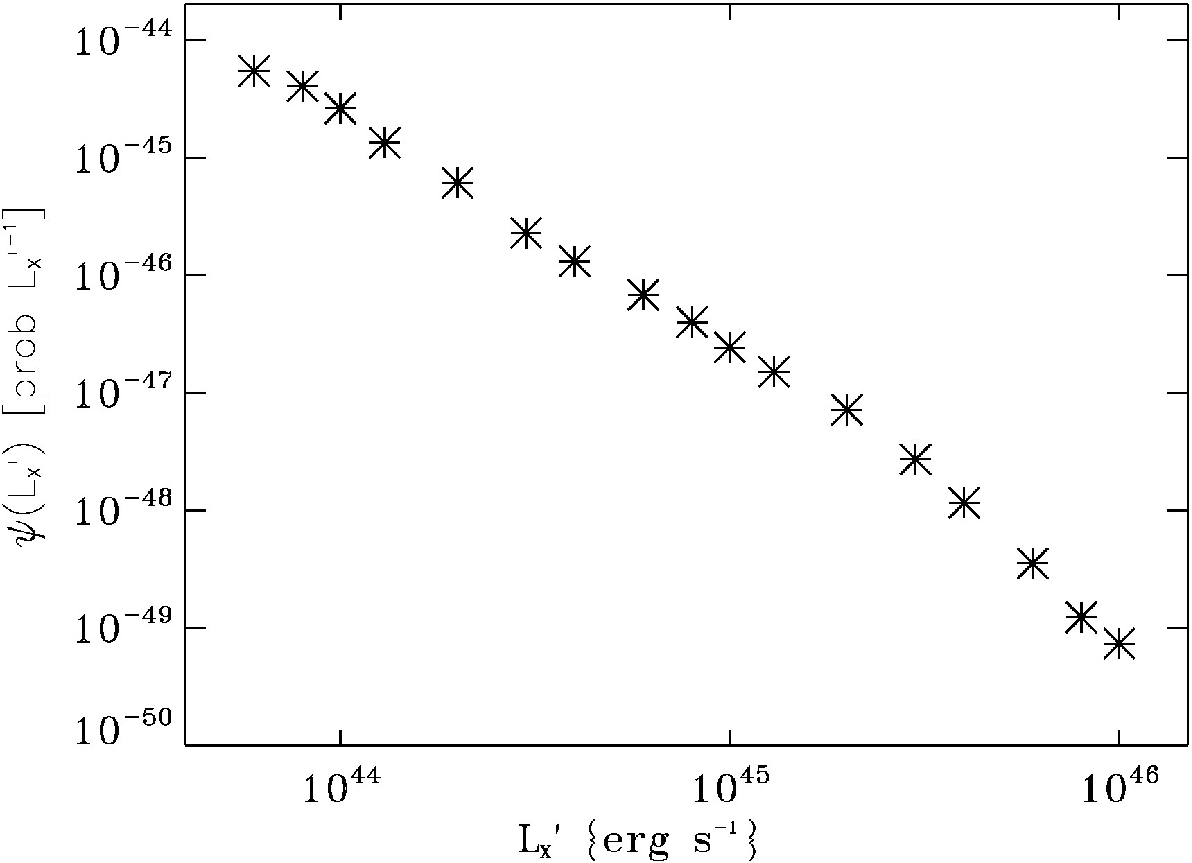}
\caption{The local X-ray LF $\psi_{\rm x}\!(L_{\rm x}')$ for quasars calculated from the sample in this work as described in \S \ref{lumsec}.  It is well fit by a broken power law with power-law slopes of -2.0$\pm$0.1 below the break and -3.1$\pm$0.1 above it.  The break occurs at around $2 \times 10^{45}$\,erg\, s$^{-1}$. }
\label{psix}
\end{figure}

Figure \ref{psix} shows the local X-ray LF.
As evident, the local X-ray LF, similar to those found in other wavebands is well-fit by a broken power law.  In this case it manifests power-law slopes of -2.0$\pm$0.1 below  and -3.1$\pm$0.1 above the  break luminosity, which is around $2 \times 10^{45}$\,erg\, s$^{-1}$.  These results can be compared to those of previous determinations of the X-ray LF of AGN such as in \citet{Ueda03}, \citet{Ueda14}, \citet{Miyaji15}, \citet{Aird15}, and \citet{Ftp16}. Table \ref{tab2} shows the results from those works for the equivalent parameters of the local LF, noting that they use an LDDE parameterization.\footnote{These works present indexes and break luminosities for $d\Phi(>L) / d\log L$ not $d\Phi(>L) /d L$, with an LDDE parameterization, so the magnitude of their power law indexes must be increased by one unit for comparing with those defined here.}  The results here are in line with these previous ones on the power law indexes, but place the break luminosity at a somewhat higher value.

\begin{table}
\scriptsize
\caption{Local luminosity function parameter results}
\label{tab2}
\begin{tabular}{cccc}
 & $\gamma_{\rm high}$ & $\gamma_{\rm low}$ & $\log L_*$ (erg s$^{-1}$) \\
\hline
this work & -3.1$\pm$0.1 & -2.0 $\pm$0.1 & 45.3$\pm$0.1 \\
\citet{Ueda03} & -3.23$\pm$0.13 & -1.86$\pm$0.15 & 43.94$^{+0.21}_{-0.26}$\\
\citet{Ueda14} & -2.71$\pm$0.09 & -1.96$\pm$0.04 &  43.97$\pm$0.06 \\
\citet{Miyaji15} & -3.80$^{+0.16}_{-0.10}$ & -2.17$\pm$0.05 & 43.97$\pm$0.06 \\
\citet{Aird15} & -3.26$\pm$0.07 & -1.72$\pm$0.02 & 44.10$\pm$0.05 \\
\citet{Ftp16} & -3.40$\pm$0.11 & -1.87$\pm$0.06 & 43.77$\pm$0.11 \\
\hline
\end{tabular}
\begin{tablenotes}
\footnotesize
\item Results for the power-law slopes above the break luminosity ($\gamma_{\rm high}$), below the break luminosity ($\gamma_{\rm low}$), and the break luminosity ($L_*$) for the local $z$=0 X-ray LF $\psi_{\rm x}\!(L_{\rm x}')$ as determined in this work and as reported by others in the literature.  
\end{tablenotes}
\end{table}

\section{Summary and Discussion} \label{disc}

We have used methods verified in \citet{CW} and previous works to derive the intrinsic correlation between the X-ray and optical-UV luminosities of quasars, utilizing the data sets discussed in \S \ref{data} and the techniques discussed in \S \ref{tech}.  As described there, much of the (raw relatively strong) observed correlation between the luminosities is induced by the observational truncations and the mutual dependencies of the two luminosities on redshift. To account for these effects we first determine the luminosity evolution in the two bands, calculate the de-evolved  (``local'') luminosities, and use partial correlation to obtain the intrinsic luminosity-luminosity correlation.  We a assume power law correlation function form between the two luminosities and using the PPCC we determine the correlation  form $L'_{\rm X} \propto ({L'_{\rm UV}})^{\gamma}$ with $\gamma=0.28\pm$0.03. 

The value of $\gamma=0.28\pm$0.03 for the correlation power law between ultraviolet and X-ray luminosities can be added to the values of 0.25$\pm$0.15 found for the optical-radio correlation and 0.75$\pm$0.10 for the optical-infrared correlation in \citet{CW} and 0.68$\pm$0.04 for the optical-gamma ray correlation reported by \citet{V21}.  These values are consistent with the canonical model of jet launching in AGN systems, where the spin energy of the central supermassive black hole is tapped to launch the jets.  We consider that the optical and ultraviolet emission arises primarily from the hot accretion disks, the infrared emission arises primarily from dusty tori surrounding accretion disks, the radio and gamma-ray emissions arise primarily from the jets, and the X-ray emission arises primarily from the accretion disk corona which can be obscured.  The high correlation seen between mid-infrared and optical luminosities in quasars lends support to the picture of tori being heated primarily by accretion disks. The significantly weaker correlation between radio and optical luminosities can be taken to support the notion that radio emission is affected by both the accretion disk size and the black hole spin, and indeed most importantly by the latter.  Now the additon of the middle-range correlation between the X-ray and optical luminosities supports a picture where the luminosity of the coronae correlate only moderately with the rest of the accretion disk.  A possible reason is that larger accretion disks may have lower temperature coronae due to the innermost stable circular orbit being located farther out for the more massive black holes that would generally have larger accretion disks \citep[e.g][]{Wilkins20}.

A surprising result of the luminosity evolution study is that the X-ray luminosities have evolved more slowly, as determined in \S \ref{techLE} and visualized in Figure \ref{alphas}, with a best-fit value of luminosity evolution index $k_{\rm x}$=0.55, while in previous works \citep{QP2,BP2,QP3} we have found best-fit values of $k_{\rm inf}$=2.5, $k_{\rm opt}$=3.3, and $k_{\rm rad}$=5.5 for quasars, and $k_{\rm \Gamma}$=5.5 for flat spectrum radio quasar-type blazars. The slow X-ray luminosity evolution, especially compared to optical-UV and infrared, which are all related to the accretion rate and the structure of the environment of the accretion disk, seems unexpected. 

Perhaps relatedly, we also find a  relatively low average empirical values for the X-ray spectral index $\varepsilon_x\sim 0.3$, as described in \S \ref{data} and visualized in Figure \ref{xks}, compared to the  canonical power law index of $\varepsilon_x$=$\sim$0.7 \citep[e.g][]{HA18}.  Both these discrepancies may indicate that absorption of the  X-ray emission (presumably from the accretion disk corona) is a significant factor in determining the observed spectrum and the total X-ray output of AGN systems. The absorption of the lower energy X-rays would tend to flatten the X-ray spectrum and reduce its flux. Then the relatively low  X-ray luminosity evolution may indicate that absorption has a redshift dependence with more absorption at higher redshifts.  Thus, under the model where the X-ray emission originates primarily from the accretion disk corona \citep[e.g][]{HA18}, X-ray luminosity would be expected to roughly track the optical / UV luminosity and track in evolution as well.  That the X-ray luminosity evolution is slower than those in other bands, in particular as compared to optical/UV, indicates that the absorbing gas is more of a factor at higher redshifts, supporting a picture where jet activity serves to clear out obscuring gas over time \citep[e.g.][]{BU12}.

Previous determinations of the X-ray luminosity evolution of AGNs with comparable parameterizations to the one here include those by \citet{Macc91}, \citet{Franc93}, \citet{Boyle93}, which have found values for similar parameterizations to the $k_{\rm x}$ factor considered here of 2.56, 2.75, and 2.8.  These works all agree that the amount of X-ray luminosity evolution is lower than that in the optical waveband but find considerably more X-ray luminosity evolution than the present work.  It should be noted, however, that these analyses relied on 420, 150, and 42 objects, respectively, and the analysis of \citet{Macc91} was limited to objects at $z<1.2$.  The sample selection biases of these datasets were not necessarily well defined, including in regard to the optical band as any quasar data set in a non-optical waveband (such as X-ray) which has spectroscopic redshift information necessarily must also have an optical observation to identify the objects as quasars and determine the redshift.   

Recently \citet{RL} reported a value for the correlation power law index between the X-ray and ultraviolet luminosities of $\gamma=0.633\pm$0.002. This $\gamma$ was calculated from the observed fluxes (or possibly luminosities) which is subject to the biases introduced because of the data truncations and redshift evolutions discussed above and in \citet{CW}.  \citet{RL} then used this observed value to arrive at a determination of the shape of the luminosity distance function in the range $1.4<z<5$ that deviates from  the  $\Lambda$CDM cosmology.  This deviation favored a larger overall matter density fraction $\Omega_m$ and an evolving dark energy equation of state.  Similar results were obtained by the same method in subsequent works using quasar samples at higher redshifts \citep{RL2}, and incorporating additional X-ray quasar catalogs \citep{RL3,RL4}. The full extent to which the complex truncation of the data used by \citet{RL} and the luminosity evolutions affect this conclusion is not clear.  In an upcoming paper \citep{CPW} we will explore the implications of these effects on determining the true value for the index $\gamma$ that must be used in such calculations, and whether an independent determination of the distance-redshift relation can be truly obtained from the observed correlations between different waveband luminosities or fluxes.

\acknowledgments

We thank Dan Wilkins for informative discussions.  Funding for the SDSS and SDSS-II has been provided by the Alfred P. Sloan Foundation, the Participating Institutions, the National Science Foundation, the U.S. Department of Energy, the National Aeronautics and Space Administration, the Japanese Monbukagakusho, the Max Planck Society, and the Higher Education Funding Council for England. The SDSS Web Site is http://www.sdss.org/.  This research has made use of data obtained from the Chandra Source Catalog, provided by the Chandra X-ray Center (CXC) as part of the Chandra Data Archive.  This research has made use of data obtained from the 4XMM XMM-Newton Serendipitous Source Catalog compiled by the 10 institutes of the XMM-Newton Survey Science Centre selected by ESA.

\end{document}